\documentstyle[preprint,aps]{revtex}
\input epsf
\begin{document}
\draft
\newcommand{\Q}{\underline{Q}}
\newcommand{\q}{{\bf q}}
\title{Phonon Absorption at the Magneto-Roton Minimum in the Fractional Quantum Hall Effect}
\author{CJ Mellor, RH Eyles, JE Digby, AJ Kent, KA Benedict, LJ Challis, M Henini, CT Foxon}
\address{Department of Physics, University of Nottingham, NOTTINGHAM, NG7 2RD, UK.}
\author{ and JJ Harris}
\address{Semiconductor Materials I.R.C., University College London, LONDON, UK.}
\date{\today}
\maketitle
\begin{abstract}
We have made the first phonon absorption measurements in the fractional 
quantum Hall r\'egime. Experiments have been conducted on two samples which have 
similar electron densities but greatly differing mobilities. The energy gaps
as measured by activation studies of the longitudinal resistance differ by a 
factor of two. Phonon absorption measurements give almost identical values for 
the energy gap demonstrating that the gap measured in this way is rather 
insensitive to disorder. The value of this gap is in agreement with the 
activation gap measured in the high mobility sample. Values obtained at 
$\nu=2/3$ are in good agreement with theory.
\end{abstract}
\pacs{Pacs: 73.40.Hm, 63.20.Kr, 73.20.Dx}

The fractional quantum Hall effect (FQHE), which occurs in a high mobility 
two-dimensional electron system (2DES) subject to a strong perpendicular 
magnetic field, is ascribed to the existence of an incompressible quantum 
liquid at certain rational filling factors. It is believed that the low lying 
excited states of the liquid are collective modes which are never gapless 
(except at the sample boundaries). Girvin, MacDonald and Platzman \cite{gmp} 
(GMP) have developed a theory for these collective modes at the primary 
filling factors $\nu=2\pi l_c^2 n_s = 1/m$ ($n_s$ is the electron sheet
density, $l_c=\sqrt{\hbar/eB}$ is the cyclotron length and $m$ is an odd 
integer) based on a modified version of Feynman's theory of the collective 
modes of superfluid Helium-4 \cite{feynman}. GMP find that the dispersion of 
the collective mode has a deep minimum for wavelengths comparable to the mean 
interparticle spacing. This `magneto-roton' minimum occurs, as in liquid 
Helium, because of a peak in the static structure factor. Recently a new (but 
presumably equivalent) framework for understanding the FQHE has emerged 
\cite{cf,hlr} based on the idea that the fractional quantum Hall effect of 
electrons is equivalent to the integer quantum Hall effect of composite 
fermions consisting of charges bound to point fluxes. The theory of the 
collective excitations within this framework is currently being developed 
\cite{cf_cm} but is not yet in quantitative agreement with the calculation of 
\cite{gmp}. What has so far been lacking is any experimental measurement
of the gap close to the magneto-roton minimum. We report on the first such 
measurements.

Experimental studies of energy gaps in the FQHE state have been pursued in
several ways including activated magneto-transport measurements where 
disorder is known to affect the values obtained \cite{disorder}. Pinczuk et al 
\cite{inelastic_light_expt} observe a feature in the inelastic light 
scattering spectrum that is attributed to the low wavevector excitations of 
the FQHE state in a quantum well. Photoluminescence experiments find anomalies in 
the spectra in the FQHE r\'egime \cite{pl_expt} but quantitative 
interpretation of these results requires a detailed understanding of the 
dynamical response of the 2DES in optical recombination processes. 

Ballistic acoustic phonons have proved to be a unique probe of low dimensional 
systems \cite{phonons}. The typical energies and wavevectors are well matched 
to those of the 2DES and since, in the FQHE state, the magneto-rotons are the 
only low energy modes which can couple to the ground state through the 
electron density, they should provide the principal channel for the absorption 
of acoustic phonons. For example, the dispersion curve of a longitudinal 
phonon incident on a 2DES in the FQHE state at an incident angle of 60 degrees 
crosses the collective mode dispersion curve close to the magneto-roton
minimum. This makes phonon absorption a promising method to investigate the 
energy gap of the FQHE in this region.

We have studied two high mobility GaAs-AlGaAs heterostructures with similar 
electron densities but greatly differing mobilities. Both devices were 
formed from heterojunctions separated by a pure AlGaAs spacer layer from a 
Silicon doped AlGaAs layer. The details of the samples are given in table 
\ref{table1}. Both devices were illuminated with red light to give the carrier 
densities quoted. A 3mm by 2mm Hall bar was defined using photolithography, 
wet chemical etching and AuGeNi contacts. The rear face of each sample was 
polished to an optical finish and $600 \mu m \times 60 \mu m$ constantan 
heaters evaporated opposite the Hall bar. The positions of the heaters were 
determined using front-to-back alignment. One heater was positioned opposite 
the middle of the Hall bar whilst the other was over an edge. 

The phonon absorption was measured from the change in longitudinal resistance 
produced by a burst of non-equilibrium ballistic phonons. The 2DES was
supplied with a constant bias current and the phonons were generated by 
applying a 100ns electrical pulse to one of the constantan heaters. The 
spectral distribution of the phonons was assumed to correspond to that of a 
black body at the temperature of the heater, $T_h$, which is calculated from 
the total power dissipated using acoustic mismatch theory \cite{mismatch} and 
confirmed to within ten per cent by measuring the energy gaps at $\nu=12$ and 
$\nu=14$ on G635. To obtain the magneto-roton gap we measured the increase in 
longitudinal resistance in the tail of the ballistic response. We assume this 
resistance change to be proportional to the absorption and have measured it as a 
function of heater temperature. The gaps deduced from the experiments were 
independent of bias current, length and repetition rate of the heater pulses 
and substrate temperature. In addition, although the magnitude of the response 
differed between the two heaters the value of the gap obtained did not.

We have also carried out measurements of the activated magnetoresistance at 
$\nu=2/3$ over a range of temperatures and fitted the results to the usual 
form $R_{xx}\sim exp\left(-\Delta_{tr}/2 k_B T\right)$ (with a correction for
hopping at low temperatures). This leads to values for 
$\Delta_{tr}$ of $2.7 \pm 0.2 K$ and $5.5 \pm 0.5 K$ for NU409 and G635 
respectively. 

The rate of energy transfer to the 2DES from the phonon beam can be expressed, 
using first order perturbation theory, as
\begin{equation}
P(T_h) = \sum_s \int \frac{d^3\Q}{{(2\pi)}^3} \hbar \omega_s(\Q) n_s(\Q) {|M_s(\Q)|}^2 S(\q,\omega_s(\Q)) \label{golden_rule}
\end{equation}
where $\Q=(\q,q_z)$ is a phonon wavevector, $\hbar\omega_s(\Q)=\hbar c_s Q$ is 
the energy of a phonon of wavevector $\Q$ and polarization $s$ 
($=LA$, $TA_1$, $TA_2$), $n_s(\Q)$ is the number of phonons in the given mode, 
$M_s(\Q)$ is the electron-phonon coupling and $S(\q,\omega)$ is the dynamic 
structure factor of the 2DES. We assume that the phonon occupation is equal to 
the Bose-Einstein distribution for all $\Q$ vectors within the geometrical 
cone subtended by the 2DES at the heater. The single mode approximation of 
\cite{gmp} amounts to assuming that the dynamic structure factor has the form
\begin{equation}
  S(\q,\omega) = \overline{s(\q)} \delta\left(\omega - \Delta(\q)\right)
\end{equation}
where $\overline{s}(\q)$ is the lowest Landau level projected {\em static} 
structure factor (obtained by GMP from the wavefunction of Laughlin 
\cite{laughlin}) and $\Delta(q)$ is the magneto-roton energy (expressed by GMP
as a functional of $\overline{s}(\q)$). We assume that the results of 
\cite{gmp} are qualitatively correct for all FQHE states, ie that $\Delta(q)$ 
has a gap for all $q$ and has a minimum at some finite value $q^*$. For 
temperatures $T_h \ll \Delta(q^*)$ we can evaluate the integral in 
equation~\ref{golden_rule} by steepest descents leading to the result 
$P(T_h) \sim n_s(\Q)={\left[exp\left(\Delta(q^*)/k_B T_h\right)-1\right]}^{-1}$.
So, if we assume that the the response, $\delta R_{xx}(T_h)$, to a heater 
pulse at constant bias current $I_b$ satisfies 
$I_b^2 \delta R_{xx}(T_h) = P(T_h)$ 
then fitting the size of the voltage pulse to the Bose distribution will yield 
$\Delta(q^*)$. The variation of the response to the phonon pulse with
inverse heater temperature for the $\nu=2/3$ fraction is shown in 
figure 1 for both samples, the lines are the predicted variation 
using the best fit value of $\Delta(q^*)$, the points are experimental data. 
Since the scale of the voltage response cannot be absolutely predicted it is 
chosen so that one experimental point lies exactly on the line (the value of 
the gap obtained is insensitive to which point is chosen).

\epsfxsize=10cm \epsfbox{fig1.eps} 

It is clear that, for all temperatures of interest, the absorption will be 
dominated by the magneto-roton minimum where the static structure factor is 
maximal, the gap is minimal and the density of magneto-roton states is 
maximal. For the primary fraction $\nu=2/3$ (equivalent to the $\nu=1/3$ 
state by particle-hole symmetry) we can use the results of \cite{gmp} to 
evaluate the integral numerically. We assume that the magneto-roton dispersion
is simply that calculated by GMP, scaled by a $q$ independent parameter which
is used as a fitting parameter. This procedure yields a value for
$\Delta(q^*)$ which is (just) within the experimental error of that 
obtained by fitting to the Bose distribution; all gaps quoted here are those 
obtained by the simpler method. One result of the more detailed calculation is 
that the primary channel for energy transfer is via the deformation potential 
coupling to the LA phonons; these are only weakly focussed by the lattice so 
we neglect phonon focusing effects entirely. More details of the calculations 
will be given elsewhere \cite{kab}. The results of these measurements are 
displayed in table~\ref{table2}.

The key point in our interpretation of the experiment is that we believe that
we are observing changes in the dissipation of the 2DES which are a direct
result of the creation of elementary excitations of the system when phonons 
are absorbed. An alternative scenario is that the 2DES is simply being heated 
by the phonon beam so that we are actually just measuring the activated 
magnetoresistance at some raised temperature $T_e$. The repetition rate of the 
heater pulse is kept low ($\sim 1kHz$) so that the average power transferred 
to the device is less than $20\mu W$, causing the substrate temperature to
rise from $50 mK$ to only $80 mK$. However the electron system could be raised 
to a higher temperature during the pulse. If the 2DES were simply being heated 
by the phonon beam then its temperature should depend only on the phonon 
intensity not on the spectral distribution. An upper bound on $T_e$ comes 
from assuming that the 2DES absorbs all of the phonons emitted by the heater, 
the temperature will then be $T_e = \alpha T_h$ where 
$\alpha={(A_h/2A_{2DES})}^{1/4}$ ($A_h$ and $A_{2DES}$ are the area of 
the heater and the 2DES respectively). For our samples the above formula gives 
$\alpha=0.23$ but in reality this is a very generous upper bound on $\alpha$ 
since not all of the ballistic phonons are incident on the 2DES and, of those 
that are, only a very small fraction are actually absorbed \cite{kent}. Hence 
our phonon measured gap, $\Delta_{\phi}$, should be related to the gap 
measured in transport experiments by 
$\Delta_{tr} = 2\alpha\Delta_{\phi} \le 0.46\Delta_{\phi}$. This inequality 
should be greatly exceeded for a realistic value of $\alpha$ but it is barely 
satisfied by the $\nu=2/3$ results on NU409 and strongly violated by the 
corresponding results for G635. Further evidence against the simple heating
interpretation comes from the fact that the magnitude of the response to the 
two heaters is quite different, indicating that the phonon intensities 
actually incident on the 2DES due to the two heaters are very different, but 
the value of the gap extracted is the same, showing that the electrical 
response depends on the spectral distribution of the phonons not their 
intensity. Hence we believe that we are seeing the direct response of the 2DES 
to the ballistic phonons.

The energy gaps determined from the phonon experiment are nearly 
equal for the two samples but those determined from activated transport 
measurements differ by a factor of just over two, in line with the difference 
in their zero field mobilities. Clearly disorder affects the two measurements 
in different ways. We interpret this difference as being due to large scale 
inhomogeneity in the lower mobility sample leading to a variation in 
the local conduction band edge. The transport measurements are taken at 
thermal equilibrium so that the measured gap will be dominated by the largest 
negative fluctuation in the band edge. Since we conclude that the phonon
experiment is a genuine local spectroscopic probe, measuring the excitation 
gap at a specific point in the 2DES, it should only be affected by short 
length scale disorder.

The energy gap obtained at $\nu=2/3$ on G635 can be compared to the 
theoretically predicted value for $\nu=1/3$. Two effects that must be 
accounted for are the finite thickness of the 2DES and Landau level mixing. In 
G635, at $\nu=2/3$, $bl_c=1.75$, where $b$ is the Fang- Howard parameter. For 
this value of $bl_c$ theoretical studies \cite{finite_width} predict that the 
energy gap is reduced by a factor of 0.7 from the ideal case of a 2DES with 
zero thickness. A rough estimate of the effect of  Landau level mixing can be 
made following the work of Yoshioka \cite{ll_mixing}. When these two effects 
are combined the theoretical gap is found to be 
$\Delta=c\times e^2/4\pi\epsilon\epsilon_0l_c^2$ where $c\approx 0.04$ in good 
agreement with our experimentally obtained value of 
$c=0.045\pm 0.002$. The small difference between the NU409 and G635 may be 
due to the differences in layer structure and doping levels; this may alter
the vertical extent of the wavefunction. Another possibility is that it is due 
to the higher disorder in NU409.  Pinzcuk et al \cite{inelastic_light_expt} 
have measured the excitation curve in the limit of low wavevector by inelastic 
light scattering in a quantum well. Allowing for the differences between the 
samples our results are in good agreement, assuming that 
$\Delta(0) \approx 2\Delta(q^*)$ \cite{gmp}.

It can be seen that the energy gaps measured by phonon absorption always lie 
above those determined by an activation measurement, even in the high mobility 
sample. As explained above we have good theoretical reasons for believing that 
the phonon absorption measurements are probing the part of the 
dispersion curve around the magneto-roton minimum. Activation measurements on 
the other hand are generally thought to measure the energy of widely separated 
quasi-electron quasi-hole pairs (hence the factor of 2 in the activated form 
of $R_{xx}$) which correspond to the large $q$ part of the collective 
excitation curve. 

The mechanism by which phonon absorption causes a change in the longitudinal
voltage is as yet unknown. One possibility is that there is a mutual friction 
between the magneto-rotons and the Laughlin liquid (possibly due to the dipole 
moment of  the former) which causes dissipation but, as pointed out by 
Platzman, \cite{platzman} this would mean that magnetotransport experiments 
measure $\Delta_{tr}=2\Delta(q^*)$ rather than $\Delta_{\infty}$. Another 
possibility is that once created by a high energy phonon, a magneto-roton can 
absorb many low energy phonons and so dissociate into an unbound quasiparticle 
and quasihole \cite{kab}. Research is in progress to determine which of these  
is the dominant effect.

There are no microscopic calculations of $\Delta(q)$ for other fractions but 
the composite fermion theory of the FQHE \cite{hlr} predicts that the energy 
gap at fractional filling factor $\nu$ is a linear function of the difference 
of the magnetic field from that at $\nu=1/2$. Our results are shown in 
figure 2. The variation of the energy gaps at $\nu=2/3$, $3/5$ and 
$4/7$ with magnetic field is consistent with the composite fermion theory of
the FQHE. The value of the composite fermion effective mass  is similar to 
that obtained  by magneto-transport \cite{cf_mass_expt}, at  
$0.6 \pm 0.1m_e$, where $m_e$ is the mass of an electron. The energy gap falls 
to zero before $\nu=1/2$ and the intercept at $\nu=1/2$ is $-0.7\pm 0.5 K$. 
The value of this intercept obtained from magneto-transport  results on the 
same sample is $\approx-2K$; this effect has previously \cite{cf_mass_expt} 
been attributed to sample disorder. Allowing for measurement uncertainties the 
effect of disorder on the absorption experiments does not exceed 0.7K. These 
observations support the argument that the absorption measurements are not 
affected by disorder as much as the magneto-transport results. 

\epsfxsize=10cm \epsfbox{fig2.eps} 
      
In conclusion the energy of the magneto-roton minimum for several filling 
factors of a high mobility heterojunction has been measured by phonon 
absorption. The value at $\nu=2/3$ is in excellent agreement with theory. The 
values for $\nu=2/3, 3/5 \mbox{ and } 4/7$ are consistent with the composite
fermion theory of the FQHE although detailed comparison with theory has not
yet been made.
      
We are grateful for support from the Engineering and Physical Sciences 
Research Council (UK) and the European Union. We would like to thank W. 
Dietsche and V. Fal'ko  for helpful discussions.

\begin{table}
\narrowtext
\caption{Properties of 2D electron systems}
\label{table1}
\begin{tabular}{||c|c|c||}
Sample & NU409 & G635\\ \hline\hline
Mobility (${cm}^2 V^{-1} s^{-1}$) & $1.0 \times {10}^6$ & $8.0 \times {10}^6$\\ \hline
Spacer layer thickness (${nm}$) &  $80$ & $60$\\ \hline
Doping Layer thickness (${nm}$) & $80$ & $200$\\ \hline
Doping level (${cm}^{-3}$) & $5 \times {10}^{17}$ & $1\times {10}^{17}$ \\ \hline
Electron Density (${cm}^{-3}$) & $1.50 \times {10}^{11}$ & $1.51 \times {10}^{11}$\\ \hline
\end{tabular}
\end{table}

\begin{table}
\narrowtext
\caption{magneto-roton Energy Gaps Measured by Phonon Absorption}
\label{table2}
\begin{tabular}{||c|c|c||}\hline
$\nu$ & $\Delta(q^*)$ (K) & $\Delta(q^*)/(e^2/4\pi\epsilon\epsilon_0l_c)$\\ \hline
$NU409$& & \\ \hline
$2/3$ & $6.2\pm 0.2$ & $0.041 \pm 0.002$\\ \hline
$G635$ & & \\ \hline
$2/3$ & $6.9 \pm 0.4$ & $0.045 \pm 0.003$\\ \hline
$3/5$ & $3.5 \pm 0.3$ & $0.021 \pm 0.002$\\ \hline
$4/7$ & $2.6 \pm 0.4$ & $0.014 \pm 0.002$\\ \hline
\end{tabular}
\end{table}
\end{document}